\documentclass[twocolumn,secnumarabic,amssymb, nobibnotes, aps, prd, showpacs]{revtex4-1}
\usepackage{latexsym, amsmath, amssymb,amsthm,amsopn,amsfonts, graphicx, epstopdf, multirow,bm}

\def\simlt{\lower.5ex\hbox{$\; \buildrel < \over \sim \;$}}
\def\simgt{\lower.5ex\hbox{$\; \buildrel > \over \sim \;$}}
\def\simgtalt{\lower.5ex\hbox{$\buildrel > \over \sim \;$}}

\def\l#1{\left #1}
\def\r#1{\right #1}

\def\eref#1{(\ref{#1})}

\begin{document}

\title{Astrophysical foregrounds  and  primordial tensor-to-scalar ratio constraints from CMB B-mode polarization observations}

\author{J. Errard}
\email{josquin, radek @apc.univ-paris-diderot.fr}
\author{R. Stompor$^*$}
\affiliation{AstroParticule et Cosmologie, Univ Paris Diderot, CNRS/IN2P3, CEA/Irfu, Obs de Paris, Sorbonne Paris Cit\'e, France }


\begin{abstract}
We study the effects of astrophysical foregrounds on the ability of Cosmic Microwave Background B-mode polarization experiments to constrain the primordial tensor-to-scalar ratio, $r$. To clean the foreground contributions we use parametric, maximum likelihood component separation technique, and consider experimental setups {\em optimized} to render a minimal level of the foreground residuals in the recovered CMB map. We consider nearly full-sky observations, include two diffuse foreground components, dust and synchrotron, and study cases with and without calibration errors,  spatial variability of the foreground properties, and partial or complete B-mode lensing signal removal.

In all these cases we find that in the limit of an arbitrarily low noise level and in the absence of the systematic effects, due to the instrument or modeling, the foreground residuals do not lead to a limit on the lowest detectable value of $r$. Nevertheless, the need to control the foreground residuals will play a major role in determining the  minimal noise levels necessary to permit a robust  detection of $r(\simlt 0.1)$ and therefore in optimizing and  forecasting the performance of the future missions.
For noise levels corresponding to current and proposed experiments, the foreground residuals are found non-negligible and potentially can significantly affect our ability to set constraints on $r$. We furthermore show how in some of these cases the constraints can be significantly improved on by restricting the post-component separation processing to a smaller sky area. This procedure applied to a case  of a COrE-like satellite mission is shown to result potentially in more than an order of magnitude improvement in the detectable value of $r$.
Our conclusions are found to be independent on the assumed overall normalization of the foregrounds and only quantitatively depend on specific parametrizations assumed for the foreground components. They however assume sufficient knowledge of the experimental bandpasses as well as foreground component scaling laws.
\end{abstract}

\maketitle

\section{Introduction}
Cosmic Microwave Background (CMB) B-mode observations are expected to be a primary source of information about the physics of the very early Universe, potentially providing an unambiguous proof of existence of the primordial gravity waves, e.g., \cite{1997PhRvL..78.2054S, 1997NewA....2..323H, 1997PhRvL..78.2058K, 1998PhRvD..57..685K}, considered a telltale signature of inflation. Consequently, the CMB B-mode observations are a very dynamic area of the current research in cosmology, with multiple observatories being designed, built, and deployed, e.g., \citep{2010SPIE.7741E..39A,2010SPIE.7741E..37R, 2011arXiv1102.2181T, 2009arXiv0903.0902A, 2011arXiv1105.2044K}. The challenges expected in carrying out such a program are as impressive as its goals exciting. Besides the observation-specific issues, related to instrumental hardware or its operations, two effects: lensing-induced B-mode signal and astrophysical foregrounds, have come to the fore, both deemed as capable of setting some ultimate limitations to the exploitation of the CMB B-mode potential \cite{2005MNRAS.360..935T, 2009A26A...503..691B}. In this paper we revisit these problems from a perspective of the parametric component separation technique~\cite{1994ApJ...424....1B, 2009MNRAS.392..216S}. Its keystone assumption is that sufficiently precise parametrizable frequency scaling laws are available for each of the relevant components. At the present, such an assumption may look somewhat farfetched, in particular, in the context of the high precision required for the B-mode data analysis. However, the parametric methods, in their modern formulation have been shown to perform very well at the present in the number of contexts, e.g., \cite{2004ApJ...612..633E, 2006ApJ...641..665E, 2009ApJS..180..306D}, and this, given the theoretical and observational work undertaken currently and aiming at understanding all the major foreground components, seems to bode well for its future applications.
In the context of this study the method provides a flexible framework, amenable to semianalytic, statistically robust analysis.

In this paper we consider experimental setups optimized to ensure the lowest foreground residual level~\cite{2011PhRvD..84f3005E}. 
We use two component foreground model, including synchrotron and dust, each parametrized with one parameter, referred to as spectral indices. We note that more refined scaling laws with potentially more parameters can be straightforwardly incorporated in the formalism presented below and would affect our conclusions only quantitatively. The foreground templates used here are described in detail in \cite{2011PhRvD..84f3005E, 2010MNRAS.408.2319S}  and we assume nearly full ($\sim 80$\%) sky coverage, corresponding to the choice of {\sc mask-i} of~\cite{2011PhRvD..84f3005E}. Given the optimized setup and its noise we estimate a typical residual and compare it with the total statistical uncertainty. As the latter depends on the tensor-to-scalar ratio, $r$, for each value of $r$ we determine respective instrumental sensitivity for which the residual is irrelevant given the uncertainties and compare it with the statistical limits due to the noise and the CMB signal only.

\section{Methodology}

\label{sect:method}

{\bf Parametric component separation:}
The fiducial data set we consider hereafter is made of multiple-single frequency maps of $Q$ and $U$ Stokes parameters, with the instrumental noise assumed to be uncorrelated, both between the pixels and
channels, pixel-independent, and characterized by its variance, $\bm{N}$. The corresponding data model we use hereafter then reads,
\begin{eqnarray}
\bm{d}_p = \bm{B}\l(\bm{\beta}, \bm{\omega}\r) \, \bm{s}_p \, + \, \bm{n}_p
\equiv \bm{\Omega}\l(\bm{\omega}\r)\,\bm{A}\l(\bm{\beta}\r) \, \bm{s}_p \, + \, \bm{n}_p,
\label{eqn:dataModelFull}
\end{eqnarray}
where for each pixel $p$, $\bm{A}$ is a mixing matrix parametrized by the spectral indices, $\bm{\beta}$, $\bm{s}_p$ -- a vector of sky signals to be recovered and $\bm{n}_p$ -- instrumental noise. $\bm{\Omega}$ is a pixel-independent, diagonal matrix with the diagonal elements, $\bm{\omega}_i \equiv \bm{\Omega}_{ii}$ 
corresponding to the calibration factors for  each of the channels. 
The likelihood function then reads~\cite{2009MNRAS.392..216S},
\begin{eqnarray}
-2 \ln{\cal L} & = & \sum_p\,\l(\bm{d}_p - \bm{B} \, \bm{s}_p\r)^t \bm{N}^{-1} \l(\bm{d}_p - \bm{B} \, \bm{s}_p\r) 
\nonumber\\
& + & 
\left[\left(\bm{\omega} - \bar{\bm{\omega}} \right)^{t} \bm{\Xi^{-1}} \left(\bm{\omega} - \bm{\bar{\omega}} \right)\right],
\label{eqn:like}
\end{eqnarray}
where the last term is simply a prior term constraining the plausible values of the calibration factors. Hereafter we will assume that the true values of the calibration factors are equal to unity, $\bar{\bm{\omega}}_i = 1$, and that their uncertainty is described by an error matrix, $\bm{\Xi}$, which for simplicity is assumed to be proportional to a unit matrix, i.e., $\bm{\Xi}_{ij} \equiv  {\sigma_{\omega}^{-2}} \, \delta_{i}^{j}$, where  $\sigma_{\omega}$ is assumed not to depend directly on the parameters of the considered experiment. 
Moreover, throughout this paper the detector's bandpasses are always taken to be known perfectly and therefore their effects on the mixing matrix, $\bm{B}$, straightforwardly calculable. The import of the bandpass uncertainties will be studied elsewhere.
In the cases without calibration uncertainty, $\bm{B} = \bm{A}$ and we simply drop the last  term. In general, we will estimate both $\bm{\beta}$ 
and $\bm{\omega}$ and maximize this likelihood to perform the component separation.\\
\\

{\bf Residual computation:}
The computation of the residuals involves two steps. First, we obtain the error of the estimation of the spectral parameters. This is done using a generalization of
Eq.~(5) of \cite{2011PhRvD..84f3005E}, allowing for the calibration errors \cite{2009MNRAS.392..216S}, and derived as the Fisher matrix, $\bm{\Sigma}_{ij} \equiv  \l.\l\langle\frac{\partial^2 \ln {\cal L}_{spec}}{\partial\gamma_i\,\partial\gamma_j}\r\rangle_{noise}\r|_{\hat{\bm{\gamma}}}^{-1}$, of the profile likelihood, ${\cal L}_{spec}$, of the likelihood in Eq.~\eref{eqn:like}, i.e., 
\begin{widetext}
\begin{eqnarray}
\bm{\Sigma}_{ij}^{-1} &  = &  n_{pix}\,\l. {\rm tr}\,\l\{\l[ \bm{B}_{,i}^t\, \bm{N}^{-1}\,  \bm{B}\, \l(\bm{B}^t\bm{N}^{-1}\bm{B}\r)^{-1} \bm{B}^t\bm{N}^{-1}\bm{B}_{,j} - \bm{B}_{,i}^t \, \bm{N}^{-1} \bm{B}_{,j}\,\r] \bm{\hat{F}}\r\} 
+   \left[\left(\bm{\omega} - \bar{\bm{\omega}} \right)^{t} \bm{\Xi^{-1}} \left(\bm{\omega} - \bm{\bar{\omega}} \right)\right]_{,ij}\r|_{\hat{\bm{\gamma}}},
 \label{eqn:secDervSat}
 \end{eqnarray}
\end{widetext}
which has to be evaluated at the true values of the parameters, $\bm{\gamma} = \bm{\hat{\gamma}}$, where $\bm{\gamma}$ stands for either $\bm{\beta}$ or $\bm{\omega}$, $_{,i} \equiv \partial/\partial\gamma_i$,  and the matrix $\bm{\hat{F}}$ defined as $\bm{\hat{F}} \equiv n_{pix}^{-1}\,\sum_p\,\bm{s}_p\,{\bm{s}_p}^t$ encapsulates all the information about the sky components needed for the parameter errors estimation. In the following
we will be removing the contribution to $\bm{\Sigma}$ related to the mode $\bm{ v} \propto [0, ..., 0, 1, ..., 1]$, where the zeros are assigned to the spectral parameters, $\bm{\beta}$, and ones to the calibration ones, $\bm{\omega}$, and $\bm{v}$ is normalized to one.  This is done by replacing $\bm{\Sigma}$ by $ \bm{\Sigma}\,-\, (\bm{v}^t\, \bm {\Sigma}\bm{v})\bm{v}\bm{v}^t$. The mode $\bm{v}$ describes an overall miscalibration of the final CMB map, RMS of which is given by $\sigma_{\bm {\omega}}$,  introducing a similar 
error in our determination of $r$, which typically is much smaller than the statistical uncertainty ($\delta r/r \simgt 0.01 \simgt \sigma_{\omega}^2$ for $r \simlt 0.1$), and thus negligible.

We use the recipe of~\cite{2010MNRAS.408.2319S} to calculate the power spectra of the typical noise-free foreground residuals, ${\bm C}_\ell^{\Delta}$,  found in the separated maps, i.e.,
\begin{eqnarray}
{\bm C}_\ell^{\Delta} \equiv
\sum_{k,k'}\sum_{j,j'}\,\bm{\Sigma}_{kk'}\,\bm{\alpha}^{0j}_k\,\bm{\alpha}^{0j'}_{k'} \, \hat{\bm{C}}^{jj'}_{\ell},
\label{eqn:resSpec0}
\end{eqnarray}
here 
$\hat{\bm{C}}^{jj'}_{\ell} $ is a cross-spectrum of components $i$ and $j$, and,
\begin{eqnarray}
	\bm{\alpha}_k \equiv \l.\frac{\partial}{\partial\, \bm{\gamma}_k}\l[
\l(\bm{B}^t\l(\bm{\gamma}\r)\bm{N}^{-1}\bm{B}\l(\bm{\gamma}\r)\r)^{-1}\bm{B}^t\l(\bm{\gamma}\r)\bm{N}^{-1}\bm{B}(\hat{\bm{\gamma}})\r]\r|_{\hat{\bm{ \gamma}}}.\nonumber
\end{eqnarray}
\\

\begin{figure*}[htbp]
   \centering
   	\includegraphics[width=17truecm]{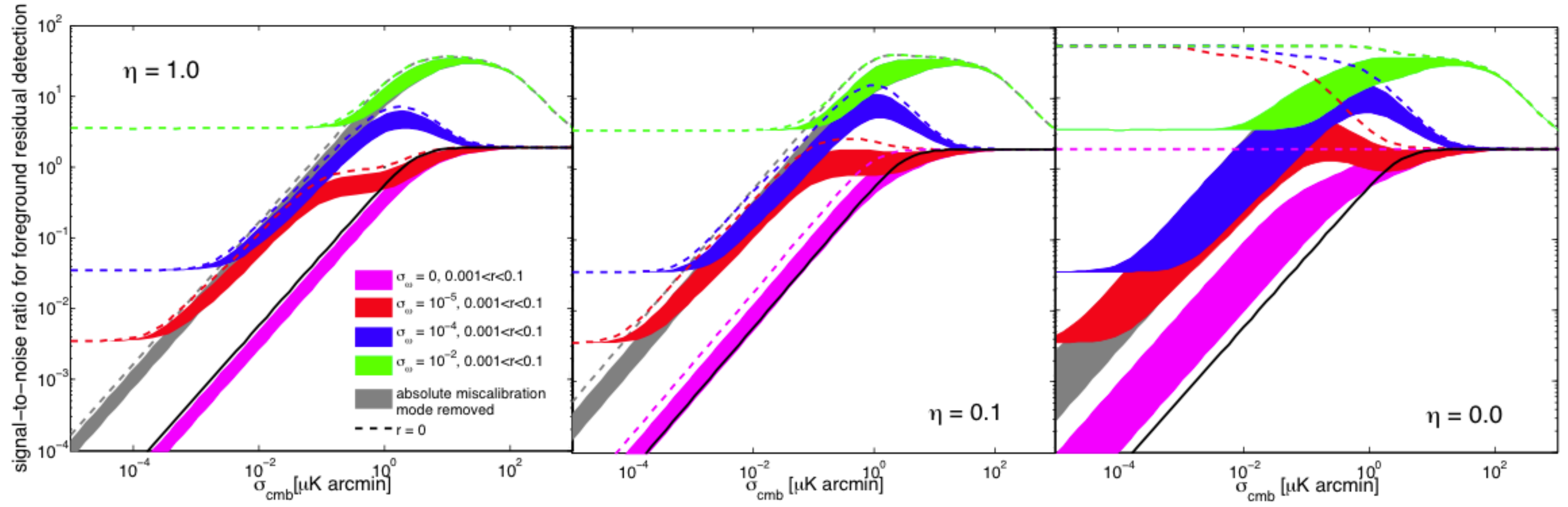}
    {\scriptsize
   \caption{\scriptsize The significance of the foreground residuals, $\sigma_{\alpha}^{-1}$, Eq.\eref{eq:exact_fisher_result}, expected in the recovered CMB map covering $\sim 80$\% of the sky for the cases with no, $\eta = 1.0$, (left panel), partial, $\eta = 0.1$, (middle panel), and complete, $\eta = 0.0$, (right panel), lensing correction, respectively. The color bands correspond to different calibration uncertainties as listed in the left panel with the gray color showing all the cases with $\sigma_\omega\ne 0$ after the removal of the mode $\bm{v}$.  The width of the shaded areas reflects the effect of varying $r$ from $0.001$, (upper edge), up to $0.1$, (lower), and the dashes show the corresponding $r=0$ cases. The black solid lines show the case with $\sigma_\omega=0$, $r=0$, and $\eta=1.0$ as a reference.}
      \label{fig:one}}
\end{figure*}

{\bf Residuals significance:}
We quantify the importance of the residuals as follows:
\begin{eqnarray}
\sigma_\alpha^{-1} = \l[f_{sky}\,\sum_{\ell}^{\ell_{max}} \,  \frac{ (2\ell + 1) C_{\ell}^{\Delta}}{C_{\ell}^{\rm prim}\l(r\r)+\,\eta\,C_{\ell}^{\rm lens}+C_\ell^{\rm noise}}\r]^{1\over2},
\label{eq:exact_fisher_result}
\end{eqnarray}
which can be derived  as a Fisher error on an overall amplitude, $\alpha(=1)$, of a foreground template, assumed to be known, with the power spectrum given by $C_{\ell}^\Delta$. $\sigma_\alpha^{-1}$ expresses statistical significance with which the template could be detected, had it been known, given the instrumental noise, $C_\ell^{\rm noise}$, and the CMB  signal, $C_{\ell}^{\rm{prim}}(r)+\eta C_{\ell}^{\rm{lens}}$. $\eta\;  (\le 1)$ denotes the fraction of the lensing signal left after its removal. 
Whenever $\sigma_\alpha^{-1}$ is large, the residual can not be neglected in an analysis of the CMB map and may need to be treated by some additional means
\cite{1475-7516-2011-08-001}. Otherwise, the foreground residuals will be irrelevant for the estimation of $r$. \\
\\
{\bf Experiment optimization:}
We use the approach described in \cite{2011PhRvD..84f3005E} to optimize the experimental setups. We assume a fixed, though arbitrary,  focal plane area during the optimization and restrict frequencies of the observational channel bands to range from $30$ GHz to $400$ GHz. The detector noise is assumed to be constant in antenna temperature units. The optimization then tries to minimize the effective $r$ value as proposed in~\cite{2007PhRvD..75h3508A}, given by $\sum_\ell^{\ell_{max}} \, C_\ell^{prim}(r_{\rm eff}) = \sum_\ell^{\ell_{max}} C_\ell^{\Delta}$. The criterion selection reflects the fact that we want to minimize the effects of the foreground residuals and thus keep their expected level as low as possible, irrespective of consequences it may have on, e.g., effective noise of the experimental configuration selected in such a way. The resulting experiment setup includes 5 frequency bands: $\nu = [30, 40, 130, 300, 400]$ GHz occupying, respectively, a fraction $f_{\rm p}(=[9, 21, 36 , 25, 9]$ per cent) of the focal plane.

\section{Discussion and results}
\label{sec:results}

Hereafter we will use the noise level of the recovered CMB map as a measure of the sensitivity of the considered experimental setups. This is given by,
\begin{eqnarray}
\sigma^2_{\rm CMB} \equiv \l[\l(\bm{B}\l(\bm{\hat{\gamma}}\r)^t\,\bm{N}^{-1}\,\bm{B}\l(\bm{\hat{\gamma}}\r)\r)^{-1}\r]_{00},
\label{eq:noise_def}
\end{eqnarray}
where we assume that CMB is the zeroth component recovered in the separation procedure. The diagonal elements of the correlation matrix, $\bm{N}$, expressing the noise level of each frequency channel, can be written in antenna temperature units as,
\begin{eqnarray}
\bm{N}_{ii} = \frac{1}{\Omega_{p}}\times\frac{4\pi\,f_{\rm sky}\,\sigma^2_{NET}}{A_{\rm fp}\, T_{\rm obs}} \times \frac{A_{\rm d}\l(\nu\l(i\r)\r)}{ \,f_{\rm fp}\l(i\r)}
\end{eqnarray}
where $\sigma_{NET}$ is a frequency-independent detector of instantaneous noise value (in $\mu {\rm K}_{\rm ant}\sqrt{\rm sec}$), $A_{\rm fp}$,  and $A_{\rm d}\l(\nu\l(i\r)\r)$ -- total  and per detector effective focal plane area, $T_{\rm obs}$ -- total observation time, and $\Omega_{\rm p}$ -- pixel size in steradians. For the considered experiment  we can write numerically,
\begin{eqnarray}
\frac{\sigma_{\rm CMB}}{\mu{\rm K}_{\rm cmb}\, \rm{arcmin}}\simeq 2.6\;10^{-3}\, \frac{\sigma_{NET}}{\mu{\rm K}_{\rm ant}}\,  \sqrt{\frac{f_{\rm sky}}{0.82}\, \frac{1{{\rm GHz}^{-2}}}{A_{\rm fp}} \, \frac{ 2  {\rm yrs}}{{T_{\rm obs}}}}.\ \ \ 
\end{eqnarray}
The dependence of our measure of the significance of the foreground residuals, $\sigma_\alpha^{-1}$, on the noise level, $\sigma_{\rm CMB}$, is illustrated in Fig.~\ref{fig:one}, and its major features can be tracked back to the behavior of the parameter errors and foreground residuals as it is discussed in detail in~\cite{ErrardInPreparation}. In particular in the low-noise regime the value of $\sigma_\alpha^{-1}$ increases $\propto \sigma_{\rm CMB}^2$ whenever no calibration uncertainty is present or the  contribution of the mode $\bm{v}$ is suppressed. This  is due to the fact that the error on all the parameters $\bm{\gamma}$ is driven by the first term on the right-hand side of Eq.~\eref{eqn:secDervSat}, resulting in a self-calibrating property of the considered system thanks to  the assumed scaling laws spanning the entire range of considered frequency bands.  The self-calibration applies only to the relative calibrations fixing the calibration coefficients of the channel maps with precision superseding that given by the assumed priors. The absolute calibration of the final map is in turn always determined by the prior term in Eq.~\eref{eqn:secDervSat} and thus independent on the experimental noise, as shown by the flat, low-noise asymptotes of the lines, computed with the mode $\bm{v}$ included. For higher noise levels the calibration errors have significant impact on the residual level and should be therefore included in any meaningful analysis.
Whenever the first term on the right-hand side of Eq.~\eref{eqn:secDervSat} is dominant,  our results also do not depend, or depend only very weakly, on the foreground amplitude and on the observed sky area (at least as long as the foregrounds are nearly stationary), as the foregrounds amplitudes present in the expressions for $\mathbf{\Sigma}$ and $C_\ell^{\Delta}$ cancel. Physically, this means that higher levels of foreground signals lead to tighter constrains on their parameters, compensating for their higher amplitudes.

\begin{figure}[htbp]
   \centering
   	\includegraphics[width=8.7truecm]{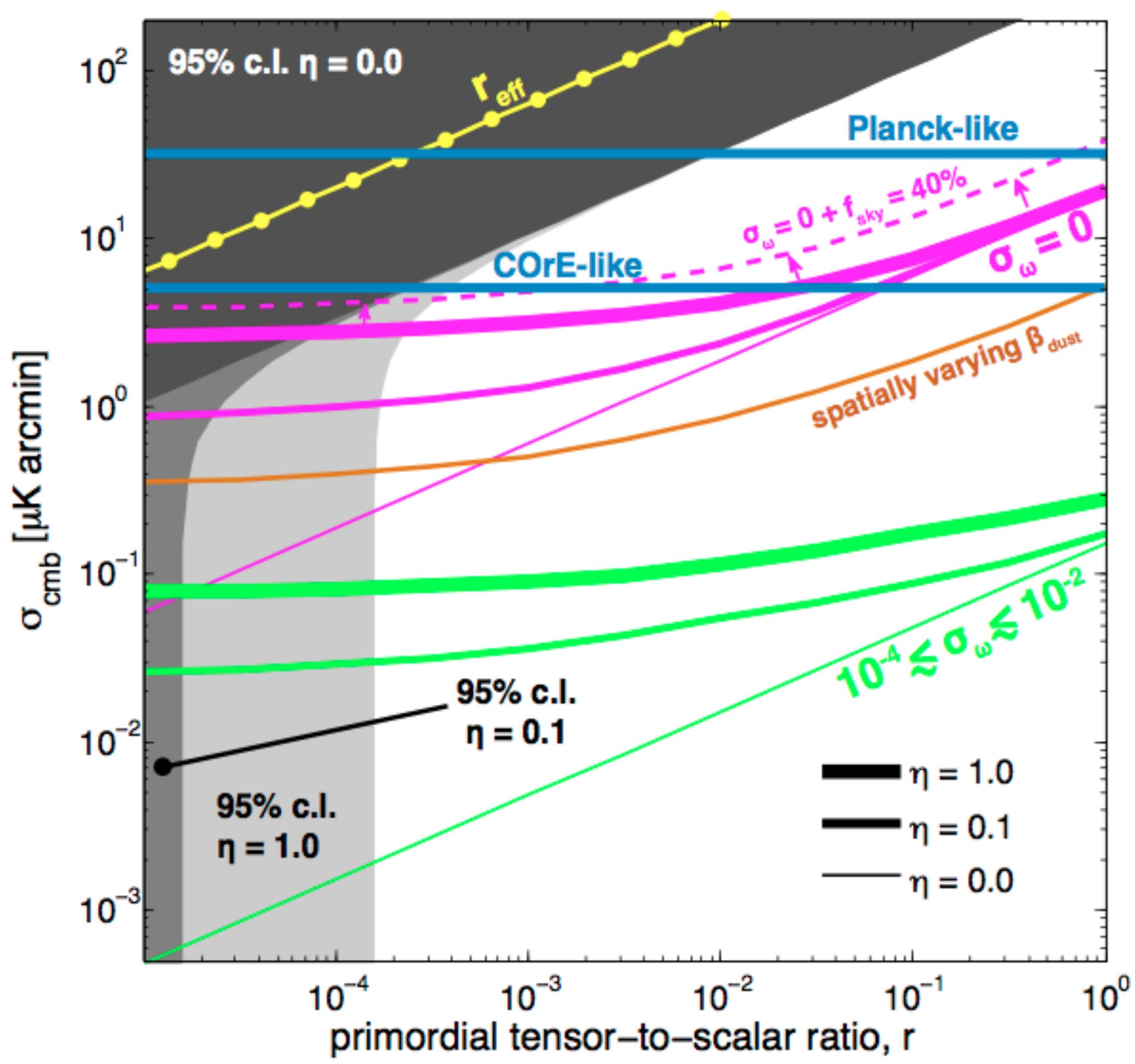} 
    {\scriptsize
   \caption{\scriptsize Upper limits on the map noise levels, which ensure that the foreground residuals are statistically irrelevant, are shown with solid lines. Each set of three lines corresponds to a different assumptions about the calibration errors as marked in the figure. In each set the lines depict the cases with no (heavy),  $90$\%  (medium), and perfect (thin) cleaning efficiency. The thin dashed line shows the change in the derived noise levels incurred as a result of restricting the sky area used to estimate $r$ {\em after} the component separation step has been already performed. These should be compared to the thick line with $\sigma_\omega=0$. The thick dots show the analogous noise limits based on an alternative criterion, $r_{eff}$, Sec.~\ref{sect:method}. The shaded areas depict statistical $2\sigma$ limits  due to the noise and sky signal for three lensing cleaning efficiencies $\eta = 1.0$, $0.1$, and $0.0$ (light to dark grey). The noise levels for re-optimized Planck and COrE-like experiments are also shown as a reference. }
      \label{fig:two}}
\end{figure}

The results from the three panels of Fig.~\ref{fig:one} are translated into limits on $\sigma_{\rm CMB}$, as shown in Fig.~\ref{fig:two},  by solving the  relation, $\sigma_{\alpha}^{-1}\l(r, \sigma_{\rm CMB}\r)=\l.\sigma_{\alpha}^{-1}\r|_{crit}$. Hereafter,  we use $\l.\sigma_{\alpha}^{-1}\r|_{crit} = 1$, corresponding to a "$1\sigma$" detection of the residuals on the map level. In general, this value should be adjusted, and the curves in the figure rescaled by $\propto  \l.\sigma_{\alpha}\r|^{-1/2}_{crit}$, given a specific application envisaged for the output maps and $1$ is used here as an illustration. 
For each $r$ value, each curve, computed for specific assumptions about the experiment and/or foregrounds, provides an upper limit on the experiments sensitivity so the foreground residuals will be found irrelevant for the analysis of the obtained CMB map.  The gray-shaded areas show the statistical uncertainties, corresponding to a different level of gravitational lensing signal cleaning. 
We note that the foreground residual limits do not prevent detecting arbitrarily low value of $r$ assuming that a
sufficiently sensitive observation can be performed. Instead, the lower limit on $r$ can arise due to a residual level of the lensing-induced B-mode signal left over from some cleaning procedure~\cite{2002PhRvL..89a1303K, 2002PhRvL..89a1304K, 2004PhRvD..69d3005S}. 
This remains true when the calibration errors are included but also when the spatial variability of the foregrounds is allowed for, and will hold at least as long as no significant deviation from the assumed component scaling laws is observed. To see the effects of the spatial variability of the spectral indices we  assume that the sky is subdivided into $n_{\rm p}$  non-overlapping patches, for each of which we assign a different set of spectral parameters. If the patches are of roughly the same size, the resulting errors on the spectral parameters will increase approximately as $\sqrt{n_{\rm p}}$, leading to a tightening of the noise constraints in Fig.~\ref{fig:two} by the same factor. For comparison the (orange) line in Fig.~\ref{fig:two}, labeled "spatially varying $\beta_{dust}$", shows a result of implementing the Stolyarov approach~\cite{2005MNRAS.357..145S}, which also leads to more restrictive noise constraints, but without introducing an ultimate limit on $r$.

We also note that by decreasing the statistical uncertainty of the map we increase $\sigma_{\alpha}^{-1}$, as the residual becomes easier to be spotted, and thus the requirements on the noise need to be tighter to ensure that the foreground level is decreased accordingly. This, for instance, explains why any lensing cleaning in Fig.~\ref{fig:two} renders a tighter limit on the noise. Conversely, re-sorting for the $r$ estimation  to a smaller map of the sky, than what has been used for the component separation, will increase the variance and lower $\sigma_{\alpha}^{-1}$, allowing us to tune appropriately  the sky area to extend the range of detectable values of $r$ given a fixed instrumental sensitivity. This will result in  lower statistical significance of the detection but will ensure that bias is negligible. This is illustrated in Fig.~\ref{fig:two} where the (magenta) arrows show a change in the noise upper  limit in the perfect calibration case, $\sigma_\omega=0$, with no lensing cleaning, $\eta=1.0$, due to using on the second step only half of the area of $80$\%  of the full sky as used for the component separation. This, for a COrE-like experiment (i.e., the proposed COrE experiment~\cite{2011arXiv1102.2181T} optimized as described earlier),  could extend its capability to detect $r$ reliably  down to $8\times 10^{-4}$ ($2\sigma$), what could be compared to $r\sim 4\times 10^{-4}$ limit ($2\sigma$) potentially achievable, if the foregrounds were absent.
We note that the trimming can be made even more efficient if the retained sky is selected to ensure the lowest possible foreground amplitude. 
If no extra trimming is done, then given our criterion for $\sigma_{\alpha}^{-1}$ the COrE-like lower limit on $r$ is found to be $r\sim 3\times 10^{-2}$,  what is at least formally within reach of a suborbital observation with similar sensitivity per pixel but observing ${\cal O}( 1)$\% of the sky~\cite{2010MNRAS.408.2319S, 1475-7516-2011-08-001}. The statistical significance of the former limit is $\sim 25\sigma$, (vs. $2\sigma$ in the suborbital case) indicating that the experimental sensitivity of such observations should be driven by the foreground separation, not  by statistical uncertainties only, but also that a further improvement of the limit on $r$ could be plausible if extra assumptions and processing are included~\cite{1475-7516-2011-08-001}. \\

The results obtained here demonstrate that in an absence of such post-component separation processing and with calibration uncertainties as typically present in actual experiments the noise levels required for an unambiguous and robust determination of $r$ are on order of ${\cal O}(10^{-1})\mu$K\,arcmin, significantly below the noise levels for the currently considered satellite mission concepts. Moreover, if the lensing contribution left over after its cleaning is higher than $\sim 10$\% of its initial value, the dependence of the noise levels on the targeted value of $r$ is rather weak. This emphasizes that once the sufficient noise level is indeed attained the measurable values of $r$ would be limited only by the statistical uncertainties. On the contrary, a failure to reach such a noise level may render the experiment incapable of setting any constraints on $r$ of current interest. If the lensing could be cleaned nearly perfectly, $\eta \simlt 10$\%, lower noise levels lead to a progressively lower limit on the detectable $r$.

Summarizing, we have studied the importance of the foreground residuals left over from the maximum likelihood parametric component separation procedure on the detection of the primordial tensor-to-scalar ratio coefficient, $r$, by nearly full-sky CMB B-mode experiments. 
We have found that though the foreground residuals are likely to be a major driver in defining the sensitivity requirements for such experiments, they do not on their own lead to any fundamental lower limits on detectable $r$, at least as long as sufficiently precise frequency scaling models are available. 
These will be rather set by the uncertainty due to the lensing signal present in the maps after its cleaning.
We note that  the latter may also in turn depend on the presence of foregrounds and instrumental noise~\cite{2002ApJ...574..566H,2009AIPC.1141..121S}, an issue we address elsewhere~\cite{ErrardInPreparation}.

\bibliographystyle{apsrev}
\bibliography{rLims}

\end{document}